\documentclass[times,twocolumn]{aastex63} 

\newcommand{\Msun}{\mbox{\,$M_\odot$}}

\newcommand{\Rsun}{\mbox{\,$R_\odot$}}

\newcommand{\ltsimeq}{\raisebox{-0.6ex}{$\,\stackrel
	{\raisebox{-.2ex}{$\textstyle <$}}{\sim}\,$}}
\newcommand{\gtsimeq}{\raisebox{-0.6ex}{$\,\stackrel
	{\raisebox{-.2ex}{$\textstyle >$}}{\sim}\,$}}

\newcommand{\EBminusV}{\mbox{$E_{B-V}$}}

\usepackage{color,soul}

\usepackage{graphics,graphicx}
\usepackage{epsfig}

\usepackage{datetime2}

\shorttitle{CO and dust in V745 Sco}
\shortauthors{Banerjee et al.}
\begin{document}
\title{Snowflakes in a furnace: formation of CO and dust in a recurrent nova eruption} 

\author[0000-0002-9670-4824]{D. P. K. Banerjee}
\affiliation{Physical Research Laboratory, Navrangpura, Ahmedabad, Gujarat 380009, India}

\author[0000-0001-6567-627X]{C. E. Woodward}
\affiliation{Minnesota Institute for Astrophysics, University of Minnesota,
116 Church Street SE, Minneapolis, MN 55455, USA}

\author[0000-0002-1457-4027]{V. Joshi}
\affiliation{Physical Research Laboratory, Navrangpura, Ahmedabad, Gujarat 380009, India}

\author[0000-0002-3142-8953]{A. Evans}
\affiliation{Astrophysics Group, Keele University, Keele, Staffordshire, ST5 5BG, UK}

\author[0000-0001-7796-1756]{F. M. Walter}
\affiliation{Department of Physics and Astronomy, Stony Brook University,
Stony Brook, NY 11794-3800} 

\author{G. H. Marion}
\affiliation{Department of Astronomy, University of Texas at Austin, 
515 Speedway, Stop C1400, Austin, Texas 78712-1205, USA}

\author[0000-0003-1039-2928]{E. Y. Hsiao}
\affiliation{Department of Physics, Florida State University, 77 Chieftan 
Way, Tallahassee, FL 32306, USA}

\author[0009-0009-1831-3914]{N. M. Ashok}
\affiliation{Physical Research Laboratory, Navrangpura, Ahmedabad, Gujarat 380009, India}

\author[0000-0003-1319-4089]{R. D. Gehrz}
\affiliation{Minnesota Institute for Astrophysics, University of Minnesota,
116 Church Street SE, Minneapolis, MN 55455, USA}

\author[0000-0002-1359-6312]{S. Starrfield}
\affiliation{School of Earth \& Space Exploration, Arizona State University, 
Box 871404, Tempe, AZ 85287-1404, USA}

\correspondingauthor{C.E. Woodward}
\email{mailto:chickw024@gmail.com}
\received{2023-06-28}
\revised{2023-08-1}
\accepted{2023-08-04}
\published{To appear in the ApJL}

\begin{abstract}
We report the detection of carbon monoxide (CO) and dust, formed under hostile conditions,  
in recurrent nova V745~Sco about 8.7 days after its 2014 outburst. The formation of 
molecules or dust has not been recorded previously in the ejecta of a recurrent nova.  
The mass and temperature of the CO and dust are estimated to be T$_{\rm{CO}} = 2250 \pm 250$~K, 
M$_{\rm{CO}} =$ (1 -- 5) $\times 10^{-8}$~\Msun\, and T$_{\rm{dust}} = 1000 \pm 50$~K ,   
M$_{\rm{dust}} \sim 10^{-8}$ -- $10^{-9}$~\Msun{} respectively. At the time of their detection, 
the shocked gas was at a high temperature of $\sim 10^{7}$~K as evidenced 
by the presence of coronal lines. The  ejecta were simultaneously irradiated by a large flux of 
soft X-ray radiation from the central white dwarf. Molecules and dust are not expected to 
form and survive in such harsh conditions; they are like snowflakes in a furnace. However, it has been 
posited in other studies that, as the nova ejecta plow through the red giant’s wind,  
a region exists between the forward and reverse shocks that is cool, dense and 
clumpy where the dust and CO could likely form. We speculate that this site may also be a region of particle 
acceleration, thereby contributing to the generation of $\gamma$-rays. 

\end{abstract}


\section{Introduction}
\label{sec-intro}

A classical nova eruption results from a thermonuclear runaway (TNR) on the surface of 
a white dwarf (WD) accreting material from a companion star in a close 
binary system. The accreted hydrogen-rich material  forms a degenerate layer on the WD’s surface 
with the mass of the layer gradually increasing with time. As the accreted matter is compressed 
and heated by the gravity of the WD, the critical temperature and pressure 
at the base of the layer reach the ignition point for the TNR that produces the nova eruption.
Observationally, the outburst is accompanied by ejection  
 of $\simeq 10^{-6}$ -- $10^{-4}$~\Msun{} of matter at high velocities and a
 large brightening, generally with an amplitude of 7 to 15 mags. In time mass-transfer from 
 the secondary resumes, leading to another eruption. All novae recur, but some do so 
 on human ($\ltsimeq 100$~yrs) timescales. To distinguish them from ``classical novae'' 
 (CNe), which have inter-eruption periods of $\simeq 10^{4}$~yrs, these are the ``recurrent novae'' 
 (RNe), defined by the selection effect that they have been seen to undergo 
 more than one explosion. 
 
 Currently there are 10 known Galactic RNe  (T Pyx, IM Nor, CI Aql, V2487 Oph, 
U Sco, V394 CrA, V745 Sco, T CrB, RS Oph, and V3890 Sgr) of which the last four belong to
a sub-class having red giant (RG) secondaries. The secondary in V745~Sco has been 
 classified as a giant implied by the luminosity class III M6$\pm$2 
\citep{1989Msngr..58...34D, 1990MNRAS.246...78S, 1991ApJ...376..721W, 1993AJ....105..320H, 1999A&A...344..177A}
with an orbital period of  2440 $\pm$ 500 days \citep{2022MNRAS.517.3640S}. 
V745~Sco is a very fast nova with $t_{2}$ and $t_{3}$ of 6.2 and 9 days respectively 
($t_{2}$, [$t_{3}$] is the elapsed time to decline 2 [3] mags from peak brightness). V745~Sco 
lies at a distance of $\simeq 8 \pm 1$~kpc in the Galactic bulge \citep{2022MNRAS.517.3640S}. 
 
Because the RG secondary in V745~Sco (and in RS Oph-type systems in general) is 
expected to  have a substantial wind, the high-velocity ejecta from the 
nova eruption plows into the RG wind creating a strong shock resulting in $\gamma$-ray and  
X-ray emission \citep{2014ATel.5879....1C, 2016ApJ...825...95D, 2015MNRAS.448L..35O, 2015MNRAS.454.3108P}, 
optical and near-infrared coronal line emission \citep[][this paper]{1989Msngr..58...34D} and 
non-thermal radio emission \citep{2015ASInC..12..107K}. The shocked gas has high temperatures, 
$10^{7} - 10^{8}$~K \citep{2014ApJ...785L..11B, 2015MNRAS.448L..35O, 2016ApJ...825...95D},
producing hard X-rays and coronal lines associated with highly-ionized atoms 
(e.g., [Fe~XIV], IP = 361~eV). 

Further, the ejecta are subjected to additional x-ray radiation from the 
central ionizing source during the supersoft phase recorded between days 3 to 10 
\citep{2015MNRAS.454.3108P, 2016ApJ...825...95D}. This is the earliest onset of the 
SSS phase in a nova. In such a harsh environment, molecules and dust cannot form or survive.  
Yet, the main result of this work shows that  carbon monoxide (CO) and dust 
\textit{did form} in V745~Sco between 8 to 10 days after outburst. 

The formation of either dust or CO in the ejecta proper of a RNe is unprecedented though 
dust features have been detected in the winds of the RG components of both RS Oph 
\citep{2007ApJ...671L.157E} and V745 Sco itself (Evans et al., in preparation). In both cases the 
features arise in dust that is a permanent feature of the binary. Here, we estimate the mass 
and temperature of the CO and dust and also investigate the formation site. 

\begin{figure}[ht!]
\figurenum{1}
\begin{center}
\includegraphics[trim=0.22cm 5.5cm 0.5cm 11.8cm, clip, width=0.57\textwidth]{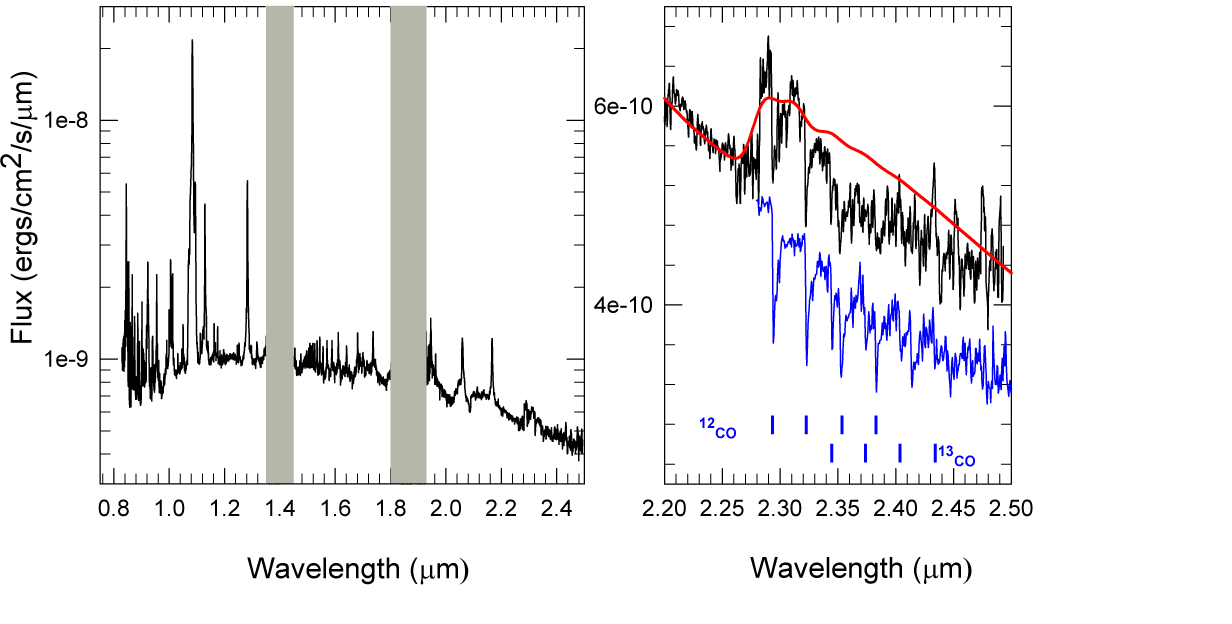}
\caption{The left panel shows the entire NIR spectrum (available as data behind the figure) of V745~Sco 8.7d after the 
2014 eruption \citep{2014ApJ...785L..11B}.  The right panel shows a magnified view 
of the first overtone CO emission (black). Superposed on the emission are the absorption band 
heads of $^{12}$CO and $^{13}$CO (vertical ticks). The spectrum of a typical M6 III, HD 18191 (blue), is 
also included to show the expected CO absorption band heads. The red line is an LTE model fit
to the CO emission. The flux axes in both panels have the same units.} 
\label{fig:fig-one}
\end{center}
\end{figure}

\vspace{1.1cm}
\section{Results}
\label{sec-results}

\subsection{Carbon monoxide in emission}
\label{sec-coemis}

Figure~\ref{fig:fig-one} shows the near infrared (NIR) spectrum 
($R \simeq 6000$, $\lambda =$ 0.8 - 2.5~\micron) of V745~Sco on 
2014 Feb 15.46 (+8.7 d after outburst) obtained using the Folded Port Infrared 
Echellette \citep[FIRE,][]{2008SPIE.7014E..0US} spectrograph on the 6.5m Magellan Baade 
Telescope. These data, presented by \citet{2014ApJ...785L..11B} and which were focused on the 
shock evolution, are re-examined here. Figure~\ref{fig:fig-one} right panel, shows a 
magnified view of the first overtone CO in emission. 
Superposed on the emission are $^{12}$CO and $^{13}$CO bandheads 
(marked in blue) which most likely  arise from the secondary with perhaps 
some contribution from neutral portions of the equatorial density enhancement 
discussed later in section~\ref{sec-co-n-dust}. 
Absorption bands are expected from the cool secondaries in RS Oph-type 
systems – it is  the emission feature that is so unexpected. CO in emission is not 
seen that often in novae. When it is detected it is always in FeII-type novae \citep{2016MNRAS.455L.109B}, 
generally forming at maximum light \citep[e.g., V705 Cas;][]{1996MNRAS.282.1049E} or  
a few weeks thereafter and being subsequently rapidly destroyed in a few days to weeks.  
It invariably precedes dust formation in CNe. Its duration and the rapidity 
of its destruction are not understood \citep{2004MNRAS.347.1294P}, but are clearly seen V2615~Oph 
\citep{2009MNRAS.398..375D} and V3662 Oph  \citep{{2017ApJ...851L..30J}}.  
\citet{2016MNRAS.455L.109B} provide a list of CO forming novae and their CO 
emission properties up until 2015.  Since 2015, four additional novae 
(V6567~Sgr, V435 CMa, V3662 Oph, V1391~Cas) have CO 
detections \citep{2017ApJ...851L..30J, 2018ATel11565....1R, 2020ATel13967....1R, 2020ATel14034....1W}. 
 
We have estimated the mass of CO using the LTE model developed 
by \citet{2009MNRAS.398..375D}, which assumes optically thin emission.  Adopting d = 8 kpc, 
the temperature and mass are found to be 2250 $\pm$ 250K, 
M$_{\rm{CO}} =$ (1 -- 5) $\times 10^{-8}$~\Msun. These are typical of values found in 
other novae and shows that in V745~Sco a substantial amount of CO was formed.  
For any other choice of distance, the CO mass scales as $(\rm{d}/8~\rm{kpc})^{2}$. The 
FWHM of each ro-vibrational line is found to be 2600 $\pm$ 300~km~s$^{-1}$ 
\citep[close to the 2200 km~s$^{-1}$ FWHM of the Pa$\beta$ line in][]{2014ApJ...785L..11B}. 
This indicates a high expansion velocity and explains the lack of band heads in the emission 
feature. It also implies that the emission originates from high velocity material i.e., from material 
associated with the shocked gas. It cannot originate from the distant unshocked RG wind 
which would typically have a small expansion velocity of 5 to 10~km~s$^{-1}$.  The CO emission is also 
seen to be blueshifted by $\simeq 2000$~km~s$^{-1}$, the reason for which is discussed in 
section~\ref{sec-co-n-dust}.


\begin{figure}[!ht]
\figurenum{2}
\begin{center}
\includegraphics[trim=7.55cm 0.15cm 0.64cm 15.8cm, clip, width=0.57\textwidth]{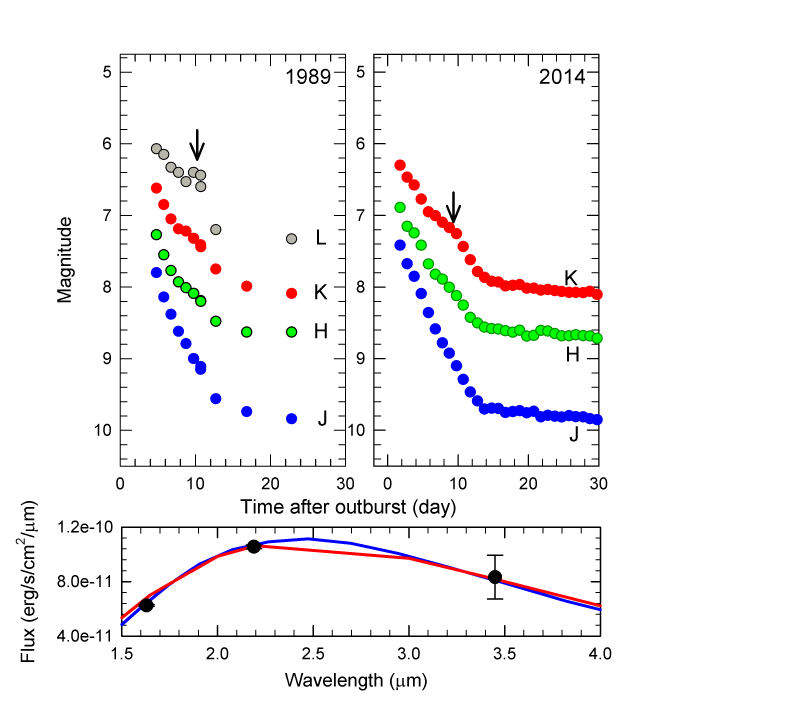}
\caption{The top two panels show the JHK(L) light curves from the 1989 
\citep{1990MNRAS.246...78S} and the 2014 eruption (ANDICAM data). The error bars 
in the plots are smaller than the symbol sizes. An IR excess 
due to dust is seen around day 9. The bottom panel shows the SED of the excess 
fitted with model fits of silicate (red) and amorphous carbon (blue) dust.  }
\label{fig:fig-two} 
\end{center}
\end{figure}

\vspace{0.95cm}
\subsection{Dust in the ejecta}
\label{sec-dust}
Dust too appears to have formed in the ejecta around the time that CO was detected. 
Figure~\ref{fig:fig-two}, upper right panel, shows high cadence JHK photometry from the
Cerro Tololo 1.3-m \citep[+ANDICAM,][]{2012PASP..124.1057W}. Between days 8 to 11, 
a clear bump is seen in the K band (marked by the arrow) and also in the H band albeit to a lesser 
extent. The bump is also present in the JHKL photometry of the 1989 eruption 
\citep{1990MNRAS.246...78S} which is shown in the upper left panel. 
Again, the IR excess is seen in the H and K bands and most prominently in the L band. 
We interpret this IR excess as emission by dust. This interpretation is consistent with 
the simultaneous appearance of CO. As mentioned earlier, all novae that have 
shown CO in emission have invariably proceeded to form dust. From the NIR light curves, 
the maximum excess over the local continuum, assumed linear in 
the region between 7 to 12 days, is measured to be $\delta$H = 0.062 $\pm$ 0.010 
and  $\delta$K = 0.155 $\pm$ 0.014 for the ANDICAM data and $\delta$L = 0.360 $\pm$ 0.2 
for the \citet{1990MNRAS.246...78S} L band data. Converting these into fluxes, after 
dereddening using \EBminusV  = 1.0 \citep{2022MNRAS.517.3640S, 2022MNRAS.517.6150S}, and assuming
that the L band excess would be the same in both 1989 and 2014, the spectral energy 
distribution (SED) of the excess based on the HKL fluxes 
is shown in the bottom panel of Figure~\ref{fig:fig-two}.

There is good  evidence for dust with the SED being well fit by models (based on the formalism 
of Banerjee et al.~2023, in press) comprising of  silicate  \citep{1984ApJ...285...89D} or 
amorphous carbon \citep[ACAR sample,][]{1996MNRAS.282.1321Z} dust grains. 
Assuming the dust emission is optically thin  and d = 8 $\pm$ 1~kpc, a dust mass and 
dust temperature of $10^{-8}$ -- $10^{-9}$~\Msun{} and $1000 \pm 50$~K  respectively 
are suggested for both silicates and amorphous carbon grains. The dust mass is within the 
range typically observed in novae \citep{1998PASP..110....3G, 2022arXiv221112410E}.
Lastly, even if the L datum is excluded from the fit in the bottom panel  because it is 
not contemporaneous with the H and K data,  the H and K points 
taken by themselves show a rising continuum indicating dust

The dust temperature is considerably  lower than  the condensation temperature  
of $\sim 2000$~K for carbon dust,  and closer to the condensation temperatures of 
silicates like forsterite, 1440~K and enstatite, 1350~K \citep{2000A&AS..146..437S}. 
The most likely condensate is therefore silicate, which might be consistent with the 
presence of silicate features in the mid-IR spectrum (Evans et al. in preparation).

The lifespan of the dust emission in V745~Sco is short with the  bump disappearing 
within a few days. Grain destruction may be by sputtering by high energy particles that diffuse 
across the shock fronts into the neutral zone or by the soft  X-ray flux 
as proposed by \citet{2017MNRAS.466.4221E} for V339~Del 
and \citet{2018ApJ...858...78G} for V5668~Sgr.

\subsection{Site of the CO and dust}
\label{sec-co-n-dust}

\begin{figure}[hb]
\figurenum{3}
\begin{center}
\includegraphics[trim=5.34cm 14.5cm 0.6cm 2.8cm, clip, width=0.49\textwidth]{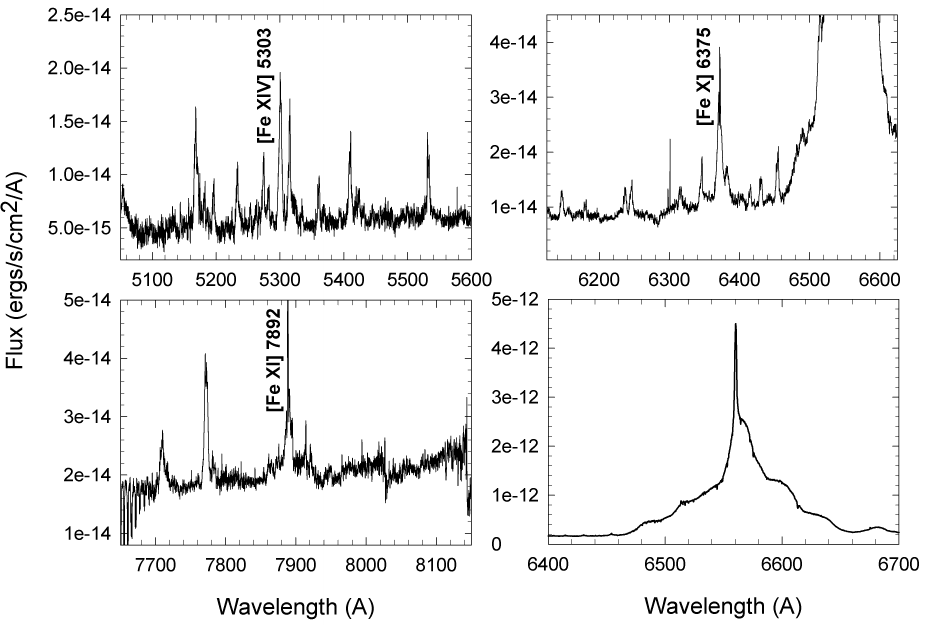}
\caption{High resolution spectra ($R = 120 000$) of V745~Sco obtained on the Cerro Tololo 1.3-m
(+CHIRON) on 2014 Feb 16 ($\sim$ +10 d past outburst) showing coronal lines from 
highly ionized species. The H$\alpha$ profile in the lower right panel is from 2014 Feb 09 
($\sim$ 3 days after the outburst) and shows structures which are discussed in the text. }
\label{fig:fig-three} 
\end{center}
\end{figure}

Between days 8.5 and 11 when CO and dust were detected, the ejecta were extremely hot 
(a few times $10^{7}$~K or more) because of shock
heating \citep{2014ApJ...785L..11B, 2015MNRAS.448L..35O, 2016ApJ...825...95D}. 
Collisional ionization is hence expected to be dominant, forming  highly ionized species. 
It is very unlikely for  molecules (CO) or dust to exist in this environment. Evidence 
for the high-ionization conditions prevailing in the ejecta on day 8.7 can be seen from 
Figure~\ref{fig:fig-three} wherein emission in several coronal lines are shown. Many 
of these ions, giving rise to the associated coronal  lines, need energies 
$\gtsimeq 350$~eV to form. In addition to the high kinetic temperature 
prevailing in the shocked gas, modeling of the supersoft x-ray emission shows
that  the black body (BB) temperature of the central ionizing source, 
after having peaked at $\sim$ 95~eV 
($1.1 \times 10^{6}$~K)  on day 5.5 -- 6.5,  was  between $7.5 \times 10^{5}$~K 
and $6 \times 10^{5}$~K on days 8 to 10 respectively \citep{2015MNRAS.454.3108P}.
Thus the temperature was very high and the radiation field was harsh, and in such an 
environment it appears unlikely that molecules could form or grains condense within the ejecta.
  
\begin{figure}[hb]
\figurenum{4}
\begin{center}
\includegraphics[width=0.40\textwidth]{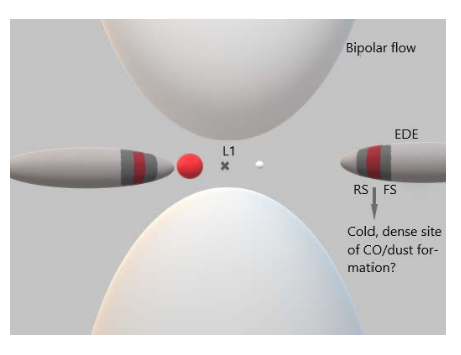}
\caption{Schematic of the V745~Sco system with the equatorial density enhancement 
(EDE) that collimates the ejecta into a bipolar flow.  The forward and reverse shocks (FS, RS)
are shown with a cold clumpy shell in between that is proposed to be the 
site of the CO and dust formation. The positions and sizes of the WD, RG and shock region 
are roughly to the following scale (section~\ref{sec-co-n-dust})
viz. R$_{\rm{RG}} = 290$~\Rsun,  R$_{\rm{WD}} = 65$~\Rsun{} 
\citep[at outburst,][]{2017MNRAS.464.5003O}, WD to RG separation = 810~\Rsun, inner 
Lagrangian point (L1) $\sim 450$~\Rsun{} from the WD. The distance of the 
FS from the WD is shown at 1070~\Rsun{} assuming  a representative 
value of the shock velocity, $v_{s} = 1000$~km~s$^{-1}$.  The actual distance lies 
at 2000~\Rsun{},  almost twice that shown. The shape of the EDE, as shown here, is only schematic. }
\label{fig:fig-four} 
\end{center}
\end{figure}

However, in the radiative shock-driven model of \citet{2017MNRAS.469.1314D} there exists a 
region where  molecules and dust could potentially form.  These authors show that 
as a shock is driven into the ejecta, molecule and dust formation can occur within the cool, dense 
shell created between the forward and reverse shocks \citep[][their Figure~1]{2017MNRAS.469.1314D}. 
While the forward and reverse shocks have temperatures of $10^{7}$~K and a few megaKelvins 
respectively, the intermediate clumpy shell is cool and dense enough due to radiative shock 
compression (particle density $\sim 10^{14}$~cm$^{-3}$) to allow CO formation and rapid dust nucleation. 
We thus propose the CO/dust seen in V745~Sco formed in this region. Based on this 
framework, Figure~\ref{fig:fig-four} shows a schematic of the V745~Sco geometry. 
This figure includes important revisions to the system spatial dimensions necessitated
by the recent revision of the orbital period \citep{2022MNRAS.517.6150S} 
which differs significantly from earlier estimates.

Although \citep{2009ApJ...697..721S} had proposed the orbital  period of V745~Sco was 
510 $\pm$ 20 days , this was shown to be incorrect by \citet{2014MNRAS.443..784M, 2015ApJS..219...26M} based 
on long term OGLE data. As a result \citet{2022MNRAS.517.3640S, 2022MNRAS.517.6150S} has re-estimated the orbital period using 
the SED method. Assuming masses of the WD (M1) and RG (M2) to be 1.39 and 1.11~\Msun{} respectively, 
\citet{2022MNRAS.517.3640S} finds the $T_{eff}^{\rm{RG}}$ and radius $R_{\rm{RG}}$ of the secondary to
be 2020 $\pm$ 60~K,  and 370 $\pm$ 50~\Rsun{} respectively and the orbital period 
to be  2440 $\pm$ 500 days.  As the estimated $T_{eff}^{\rm{RG}}$ appeared to be on 
the lower side for a  M6~III secondary, we have independently redone the analysis. 

We constructed a SED using V, I magnitudes from OGLE \citep{2015ApJS..219...26M}
as listed in SIMBAD, JHK magnitudes from 2MASS \citep{2006AJ....131.1163S}, 
and WISE \citep{2010AJ....140.1868W} W1, W2 magnitudes (\EBminusV  = 1.0 was used 
for dereddening). The WISE W3 and W4 magnitudes were excluded from the 
SED fit as they are heavily affected by dust in the V745~Sco system as shown by 
Spitzer spectra (Evans et al.~in preparation). The SED analysis  
suggests $T_{eff}^{\rm{RG}} $ = 2350 $\pm$ 50~K with the corresponding radius 
of the donor being 290 $\pm$ 44~\Rsun{} (assuming d = 8 $\pm$ 1~kpc). Assuming 
the star fills its Roche lobe, the orbital period is $\sim$ 1700~days, and orbital 
separation is $\sim$  810~\Rsun. M1 and M2 are taken from \citet{2022MNRAS.517.3640S}.
We adopt these parameters in our analysis. These estimates of the orbital period and 
separation are marginally smaller (20\% to 30\%) than the \citet{2022MNRAS.517.3640S, 2022MNRAS.517.6150S} estimates.

The spatial dimensions estimated above are used to prepare Figure~\ref{fig:fig-four} which 
is more or less to scale.  Our primary focus is to locate the position of the clumpy 
CO and dust emitting zone with respect to the RG (i.e., did the CO form before or after the blast 
wave encountered the RG?).  Figure~3 of \citet{2014ApJ...785L..11B} shows 
the decline of the FWHM of the Pa$\beta$ line with time. Taking the expansion 
velocity of the shock V(exp) as equal to half the FWHM, 
we find from the data of \citet{2014ApJ...785L..11B} that V(exp) 
can be well represented by the polynomial V(exp)  = v$_{0}$ +  at + bt$^{2}$ + ct$^{3}$, with ``t''  
being the time (in days) after outburst and with v$_{0}$ = 2689~km~s$^{-1}$, 
a = -229,  b = 5.45 and c = 0.075, respectively.  Using the above polynomial, 
it is found the shock reaches the RG’s WD-facing surface within  2 to  2.2 days. So it is obvious 
that the CO and dust, which formed between 8 to 10 days,  did not form in the 
intervening gap between the WD and RG.  In fact, at 8.7~d when the CO is detected, the 
CO emission zone was located  $\sim$ 2000~\Rsun{} away from the WD. 
That is, the shock had reached and engulfed the RG, refracted around it 
and was located  at a distance of $\sim 2000$~\Rsun{} (9.3~au) 
from the WD when the dust/CO were detected.  

It is moot to ask whether or not there is anything specific about the physical conditions 
(gas density, gas temperature, etc.)  at this  site/position that is specially favorable for 
dust/CO formation. However, it is essential to have newer 3D simulations that use the 
revised values  for the orbital separation and radius of the donor instead of the old (but incorrect) 
values used by \citet{2017MNRAS.464.5003O} which were: secondary star radius = 126~\Rsun; binary 
separation $a$ = 1.7~au = 364~\Rsun. It would be of particular interest to see what the 
revised 3D simulations show and whether they  can reproduce a clumpy, cool zone in between the 
forward shock (FS) and the reverse shock (RS) as predicted by \citep{2017MNRAS.469.1314D}. 


\section{Discussion}
\label{sec-disc}
This study presents a rare example of  a nova in which dust and CO formation are likely 
triggered by shocks \pagebreak 


\noindent \citep[another example of a shock-induced dust forming nova is  
V2891 Cyg,][]{2022MNRAS.510.4265K}  This study is also the first to record  dust and 
CO formation in a RN. With typical recurrence timescales of a few tens of years 
and assuming a high mass accretion rate of $10^{-8}$\Msun{} yr$^{-1}$, not much 
more than $10^{-7}$~\Msun{} can be accreted and subsequently ejected in a 
RN nova eruption. The ejecta are hence unlikely to be dense enough to be 
conducive to grain nucleation or molecule formation. It is however difficult to 
understand why similar systems (RS Oph, V407~Cyg, V3890~Sgr) failed to form 
CO and dust. All three objects have high cadence JHK spectra  from the start of their 
respective outbursts;  the first two on an almost daily 
basis \citep{2006ApJ...653L.141D, 2009MNRAS.399..357B, 2022MNRAS.517.6077E}. 

Among other plausible explanations, the blue-shifted CO
emission seen in V745~Sco could be a consequence of 
the equatorial density enhancement (EDE) being viewed edge-on thereby hiding the 
receding red component from view. This needs to be verified through morpho-kinematic
modeling \citep[e.g., SHAPE,][]{2011apn5.confP..59S} of high-resolution line profiles, 
for example the H$\alpha$ profile in Figure~\ref{fig:fig-three}.  The shoulders in 
the H$\alpha$ profile at $\sim$ 6480 and 6640~\AA{} profile indicative a 
collimated polar flow and therefore give observational support for an EDE. 

This study's robust evidence for a cool compressed zone 
between the forward shock (FS) and reverse shock (RS) suggests that the 
magnetic field lines from the RG should get compressed in this region due to 
the compression of matter \citep[assuming the lines of force are frozen in;][]{2007ApJ...663L.101T}.
Future exploration is necessary to determine if this site is a region of particle acceleration 
(along the lines proposed for RS~Oph in \citet{2007ApJ...663L.101T, 2023arXiv230201276T}), 
thereby contributing to the generation of $\gamma$-rays.  

\section{Summary}
\label{sec-sum}

We have reanalyzed spectra and infrared light curve data of the recurrent nova V745~Sco
shortly after its 2014 outburst. Carbon monoxide (CO) and dust in the ejecta of  recurrent nova 
V745~Sco are reported. This is a rare instance of what appears to be shock-induced dust 
formation in a nova. The mass and temperature of the CO and dust are estimated to be 
T$_{\rm{CO}} = 2250$ $\pm$ 250~K, M$_{\rm{CO}} = (1 - 5) \times 10^{-8}$~\Msun{}
and T$_{\rm{dust}} = 1000 \pm 50$~K ,  M$_{\rm{dust}} \sim 10^{-8}$ -- $10^{-9}$~\Msun{}, 
respectively. Lastly, the site of the CO/dust emission, when it was first detected, is estimated to 
lie at $\sim 2000$~\Rsun{} (9.3~au) from the white dwarf. In comparison, 
the binary separation is 810~\Rsun. 
 
\vspace{0.1cm}
\begin{acknowledgments}

The authors wish to thank the referee for their insightful comments and detailed suggestions that
improved the manuscript.
This publication makes use of data products from the Wide-field Infrared Survey 
Explorer (\url{https://doi: 10.26131/IRSA142}), which is a joint project of the University of California, 
Los Angeles, and the Jet Propulsion Laboratory/California Institute of Technology, funded by 
the National Aeronautics and Space Administration.  This publication makes 
use of data products from the Two Micron All Sky Survey (\url{https://doi: 10.26131/IRSA2}), which is a 
joint project of the University of Massachusetts and the Infrared Processing and Analysis 
Center/California Institute of Technology, funded by the National Aeronautics and Space Administration 
and the National Science Foundation. This publication, in part, is based on observations at 
Cerro Tololo Inter-American Observatory at NSF’s NOIRLab, which is managed by the 
Association of Universities for Research in Astronomy (AURA) under a cooperative 
agreement with the National Science Foundation. 

\end{acknowledgments}


\facilities{2MASS, OGLE, WISE, CTIO:1.3m, ANDICAM, CHIRON}

\software{Astropy \citep{2018AJ....156..123A} }

\clearpage


\bibliography{test3}{}

\begin{thebibliography}{}
\expandafter\ifx\csname natexlab\endcsname\relax\def\natexlab#1{#1}\fi
\providecommand{\url}[1]{\href{#1}{#1}}
\providecommand{\dodoi}[1]{doi:~\href{http://doi.org/#1}{\nolinkurl{#1}}}
\providecommand{\doeprint}[1]{\href{http://ascl.net/#1}{\nolinkurl{http://ascl.net/#1}}}
\providecommand{\doarXiv}[1]{\href{https://arxiv.org/abs/#1}{\nolinkurl{https://arxiv.org/abs/#1}}}

\bibitem[{{Anupama} \& {Miko{\l}ajewska}(1999)}]{1999A&A...344..177A}
{Anupama}, G.~C., \& {Miko{\l}ajewska}, J. 1999, \aap, 344, 177,
  \dodoi{10.48550/arXiv.astro-ph/9812432}

\bibitem[{{Astropy Collaboration} {et~al.}(2018){Astropy Collaboration},
  {Price-Whelan}, {Sip{\H{o}}cz}, {G{\"u}nther}, {Lim}, {Crawford}, {Conseil},
  {Shupe}, {Craig}, {Dencheva}, {Ginsburg}, {VanderPlas}, {Bradley},
  {P{\'e}rez-Su{\'a}rez}, {de Val-Borro}, {Aldcroft}, {Cruz}, {Robitaille},
  {Tollerud}, {Ardelean}, {Babej}, {Bach}, {Bachetti}, {Bakanov}, {Bamford},
  {Barentsen}, {Barmby}, {Baumbach}, {Berry}, {Biscani}, {Boquien}, {Bostroem},
  {Bouma}, {Brammer}, {Bray}, {Breytenbach}, {Buddelmeijer}, {Burke},
  {Calderone}, {Cano Rodr{\'\i}guez}, {Cara}, {Cardoso}, {Cheedella}, {Copin},
  {Corrales}, {Crichton}, {D'Avella}, {Deil}, {Depagne}, {Dietrich}, {Donath},
  {Droettboom}, {Earl}, {Erben}, {Fabbro}, {Ferreira}, {Finethy}, {Fox},
  {Garrison}, {Gibbons}, {Goldstein}, {Gommers}, {Greco}, {Greenfield},
  {Groener}, {Grollier}, {Hagen}, {Hirst}, {Homeier}, {Horton}, {Hosseinzadeh},
  {Hu}, {Hunkeler}, {Ivezi{\'c}}, {Jain}, {Jenness}, {Kanarek}, {Kendrew},
  {Kern}, {Kerzendorf}, {Khvalko}, {King}, {Kirkby}, {Kulkarni}, {Kumar},
  {Lee}, {Lenz}, {Littlefair}, {Ma}, {Macleod}, {Mastropietro}, {McCully},
  {Montagnac}, {Morris}, {Mueller}, {Mumford}, {Muna}, {Murphy}, {Nelson},
  {Nguyen}, {Ninan}, {N{\"o}the}, {Ogaz}, {Oh}, {Parejko}, {Parley}, {Pascual},
  {Patil}, {Patil}, {Plunkett}, {Prochaska}, {Rastogi}, {Reddy Janga},
  {Sabater}, {Sakurikar}, {Seifert}, {Sherbert}, {Sherwood-Taylor}, {Shih},
  {Sick}, {Silbiger}, {Singanamalla}, {Singer}, {Sladen}, {Sooley},
  {Sornarajah}, {Streicher}, {Teuben}, {Thomas}, {Tremblay}, {Turner},
  {Terr{\'o}n}, {van Kerkwijk}, {de la Vega}, {Watkins}, {Weaver}, {Whitmore},
  {Woillez}, {Zabalza}, \& {Astropy Contributors}}]{2018AJ....156..123A}
{Astropy Collaboration}, {Price-Whelan}, A.~M., {Sip{\H{o}}cz}, B.~M., {et~al.}
  2018, \aj, 156, 123, \dodoi{10.3847/1538-3881/aabc4f}

\bibitem[{{Banerjee} {et~al.}(2009){Banerjee}, {Das}, \&
  {Ashok}}]{2009MNRAS.399..357B}
{Banerjee}, D.~P.~K., {Das}, R.~K., \& {Ashok}, N.~M. 2009, \mnras, 399, 357,
  \dodoi{10.1111/j.1365-2966.2009.15279.x}

\bibitem[{{Banerjee} {et~al.}(2014){Banerjee}, {Joshi}, {Venkataraman},
  {Ashok}, {Marion}, {Hsiao}, \& {Raj}}]{2014ApJ...785L..11B}
{Banerjee}, D.~P.~K., {Joshi}, V., {Venkataraman}, V., {et~al.} 2014, \apjl,
  785, L11, \dodoi{10.1088/2041-8205/785/1/L11}

\bibitem[{{Banerjee} {et~al.}(2016){Banerjee}, {Srivastava}, {Ashok}, \&
  {Venkataraman}}]{2016MNRAS.455L.109B}
{Banerjee}, D.~P.~K., {Srivastava}, M.~K., {Ashok}, N.~M., \& {Venkataraman},
  V. 2016, \mnras, 455, L109, \dodoi{10.1093/mnrasl/slv163}

\bibitem[{{Cheung} {et~al.}(2014){Cheung}, {Jean}, \&
  {Shore}}]{2014ATel.5879....1C}
{Cheung}, C.~C., {Jean}, P., \& {Shore}, S.~N. 2014, The Astronomer's Telegram,
  5879, 1

\bibitem[{{Das} {et~al.}(2006){Das}, {Banerjee}, \&
  {Ashok}}]{2006ApJ...653L.141D}
{Das}, R., {Banerjee}, D. P.~K., \& {Ashok}, N.~M. 2006, \apjl, 653, L141,
  \dodoi{10.1086/510674}

\bibitem[{{Das} {et~al.}(2009){Das}, {Banerjee}, \&
  {Ashok}}]{2009MNRAS.398..375D}
{Das}, R.~K., {Banerjee}, D.~P.~K., \& {Ashok}, N.~M. 2009, \mnras, 398, 375,
  \dodoi{10.1111/j.1365-2966.2009.15141.x}

\bibitem[{{Derdzinski} {et~al.}(2017){Derdzinski}, {Metzger}, \&
  {Lazzati}}]{2017MNRAS.469.1314D}
{Derdzinski}, A.~M., {Metzger}, B.~D., \& {Lazzati}, D. 2017, \mnras, 469,
  1314, \dodoi{10.1093/mnras/stx829}

\bibitem[{{Draine} \& {Lee}(1984)}]{1984ApJ...285...89D}
{Draine}, B.~T., \& {Lee}, H.~M. 1984, \apj, 285, 89, \dodoi{10.1086/162480}

\bibitem[{{Drake} {et~al.}(2016){Drake}, {Delgado}, {Laming}, {Starrfield},
  {Kashyap}, {Orlando}, {Page}, {Hernanz}, {Ness}, {Gehrz}, {van Rossum}, \&
  {Woodward}}]{2016ApJ...825...95D}
{Drake}, J.~J., {Delgado}, L., {Laming}, J.~M., {et~al.} 2016, \apj, 825, 95,
  \dodoi{10.3847/0004-637X/825/2/95}

\bibitem[{{Duerbeck}(1989)}]{1989Msngr..58...34D}
{Duerbeck}, H.~W. 1989, The Messenger, 58, 34

\bibitem[{{Evans} {et~al.}(1996){Evans}, {Geballe}, {Rawlings}, \&
  {Scott}}]{1996MNRAS.282.1049E}
{Evans}, A., {Geballe}, T.~R., {Rawlings}, J.~M.~C., \& {Scott}, A.~D. 1996,
  \mnras, 282, 1049, \dodoi{10.1093/mnras/282.3.1049}

\bibitem[{{Evans} {et~al.}(2022){Evans}, {Geballe}, {Woodward}, {Banerjee},
  {Gehrz}, {Starrfield}, \& {Shahbandeh}}]{2022MNRAS.517.6077E}
{Evans}, A., {Geballe}, T.~R., {Woodward}, C.~E., {et~al.} 2022, \mnras, 517,
  6077, \dodoi{10.1093/mnras/stac2363}

\bibitem[{{Evans} \& {Gehrz}(2022)}]{2022arXiv221112410E}
{Evans}, A., \& {Gehrz}, R.~D. 2022, arXiv e-prints, arXiv:2211.12410,
  \dodoi{10.48550/arXiv.2211.12410}

\bibitem[{{Evans} {et~al.}(2007){Evans}, {Woodward}, {Helton}, {van Loon},
  {Barry}, {Bode}, {Davis}, {Drake}, {Eyres}, {Geballe}, {Gehrz}, {Kerr},
  {Krautter}, {Lynch}, {Ness}, {O'Brien}, {Osborne}, {Page}, {Rudy}, {Russell},
  {Schwarz}, {Starrfield}, \& {Tyne}}]{2007ApJ...671L.157E}
{Evans}, A., {Woodward}, C.~E., {Helton}, L.~A., {et~al.} 2007, \apjl, 671,
  L157, \dodoi{10.1086/524944}

\bibitem[{{Evans} {et~al.}(2017){Evans}, {Banerjee}, {Gehrz}, {Joshi}, {Ashok},
  {Ribeiro}, {Darnley}, {Woodward}, {Sand}, {Marion}, {Diamond}, {Eyres},
  {Wagner}, {Helton}, {Starrfield}, {Shenoy}, {Krautter}, {Vacca}, \&
  {Rushton}}]{2017MNRAS.466.4221E}
{Evans}, A., {Banerjee}, D.~P.~K., {Gehrz}, R.~D., {et~al.} 2017, \mnras, 466,
  4221, \dodoi{10.1093/mnras/stw3334}

\bibitem[{{Gehrz} {et~al.}(1998){Gehrz}, {Truran}, {Williams}, \&
  {Starrfield}}]{1998PASP..110....3G}
{Gehrz}, R.~D., {Truran}, J.~W., {Williams}, R.~E., \& {Starrfield}, S. 1998,
  \pasp, 110, 3, \dodoi{10.1086/316107}

\bibitem[{{Gehrz} {et~al.}(2018){Gehrz}, {Evans}, {Woodward}, {Helton},
  {Banerjee}, {Srivastava}, {Ashok}, {Joshi}, {Eyres}, {Krautter}, {Kuin},
  {Page}, {Osborne}, {Schwarz}, {Shenoy}, {Shore}, {Starrfield}, \&
  {Wagner}}]{2018ApJ...858...78G}
{Gehrz}, R.~D., {Evans}, A., {Woodward}, C.~E., {et~al.} 2018, \apj, 858, 78,
  \dodoi{10.3847/1538-4357/aaba81}

\bibitem[{{Harrison} {et~al.}(1993){Harrison}, {Johnson}, \&
  {Spyromilio}}]{1993AJ....105..320H}
{Harrison}, T.~E., {Johnson}, J.~J., \& {Spyromilio}, J. 1993, \aj, 105, 320,
  \dodoi{10.1086/116429}

\bibitem[{{Joshi} {et~al.}(2017){Joshi}, {Banerjee}, \&
  {Srivastava}}]{2017ApJ...851L..30J}
{Joshi}, V., {Banerjee}, D.~P.~K., \& {Srivastava}, M. 2017, \apjl, 851, L30,
  \dodoi{10.3847/2041-8213/aa9d86}

\bibitem[{{Kantharia} {et~al.}(2015){Kantharia}, {Dutta}, {Roy}, {Anupama},
  {Chitale}, {Ishwara-Chandra}, {Prabhu}, {Ashok}, \&
  {Banerjee}}]{2015ASInC..12..107K}
{Kantharia}, N.~G., {Dutta}, P., {Roy}, N., {et~al.} 2015, in Astronomical
  Society of India Conference Series, Vol.~12, Astronomical Society of India
  Conference Series, 107--108, \dodoi{10.48550/arXiv.1510.01120}

\bibitem[{{Kumar} {et~al.}(2022){Kumar}, {Srivastava}, {Banerjee}, {Woodward},
  {Munari}, {Evans}, {Joshi}, {Dallaporta}, \& {Page}}]{2022MNRAS.510.4265K}
{Kumar}, V., {Srivastava}, M.~K., {Banerjee}, D. P.~K., {et~al.} 2022, \mnras,
  510, 4265, \dodoi{10.1093/mnras/stab3772}

\bibitem[{{Mr{\'o}z} {et~al.}(2014){Mr{\'o}z}, {Poleski}, {Udalski},
  {Soszy{\'n}ski}, {Szyma{\'n}ski}, {Kubiak}, {Pietrzy{\'n}ski}, {Wyrzykowski},
  {Ulaczyk}, {Koz{\l}owski}, {Pietrukowicz}, \&
  {Skowron}}]{2014MNRAS.443..784M}
{Mr{\'o}z}, P., {Poleski}, R., {Udalski}, A., {et~al.} 2014, \mnras, 443, 784,
  \dodoi{10.1093/mnras/stu1181}

\bibitem[{{Mr{\'o}z} {et~al.}(2015){Mr{\'o}z}, {Udalski}, {Poleski},
  {Soszy{\'n}ski}, {Szyma{\'n}ski}, {Pietrzy{\'n}ski}, {Wyrzykowski},
  {Ulaczyk}, {Koz{\l}owski}, {Pietrukowicz}, \&
  {Skowron}}]{2015ApJS..219...26M}
{Mr{\'o}z}, P., {Udalski}, A., {Poleski}, R., {et~al.} 2015, \apjs, 219, 26,
  \dodoi{10.1088/0067-0049/219/2/26}

\bibitem[{{Orio} {et~al.}(2015){Orio}, {Rana}, {Page}, {Sokoloski}, \&
  {Harrison}}]{2015MNRAS.448L..35O}
{Orio}, M., {Rana}, V., {Page}, K.~L., {Sokoloski}, J., \& {Harrison}, F. 2015,
  \mnras, 448, L35, \dodoi{10.1093/mnrasl/slu195}

\bibitem[{{Orlando} {et~al.}(2017){Orlando}, {Drake}, \&
  {Miceli}}]{2017MNRAS.464.5003O}
{Orlando}, S., {Drake}, J.~J., \& {Miceli}, M. 2017, \mnras, 464, 5003,
  \dodoi{10.1093/mnras/stw2718}

\bibitem[{{Page} {et~al.}(2015){Page}, {Osborne}, {Kuin}, {Henze}, {Walter},
  {Beardmore}, {Bode}, {Darnley}, {Delgado}, {Drake}, {Hernanz}, {Mukai},
  {Nelson}, {Ness}, {Schwarz}, {Shore}, {Starrfield}, \&
  {Woodward}}]{2015MNRAS.454.3108P}
{Page}, K.~L., {Osborne}, J.~P., {Kuin}, N.~P.~M., {et~al.} 2015, \mnras, 454,
  3108, \dodoi{10.1093/mnras/stv2144}

\bibitem[{{Pontefract} \& {Rawlings}(2004)}]{2004MNRAS.347.1294P}
{Pontefract}, M., \& {Rawlings}, J.~M.~C. 2004, \mnras, 347, 1294,
  \dodoi{10.1111/j.1365-2966.2004.07330.x}

\bibitem[{{Rudy} {et~al.}(2018){Rudy}, {Mauerhan}, {Crawford}, {Russell}, \&
  {Wiktorowicz}}]{2018ATel11565....1R}
{Rudy}, R., {Mauerhan}, J., {Crawford}, K., {Russell}, R., \& {Wiktorowicz}, S.
  2018, The Astronomer's Telegram, 11565, 1

\bibitem[{{Russell} {et~al.}(2020){Russell}, {Sitko}, {Rudy}, {Fujii}, {Arai},
  \& {Kawakita}}]{2020ATel13967....1R}
{Russell}, R.~W., {Sitko}, M.~L., {Rudy}, R.~J., {et~al.} 2020, The
  Astronomer's Telegram, 13967, 1

\bibitem[{{Schaefer}(2009)}]{2009ApJ...697..721S}
{Schaefer}, B.~E. 2009, \apj, 697, 721, \dodoi{10.1088/0004-637X/697/1/721}

\bibitem[{{Schaefer}(2022{\natexlab{a}})}]{2022MNRAS.517.3640S}
---. 2022{\natexlab{a}}, \mnras, 517, 3640, \dodoi{10.1093/mnras/stac2089}

\bibitem[{{Schaefer}(2022{\natexlab{b}})}]{2022MNRAS.517.6150S}
---. 2022{\natexlab{b}}, \mnras, 517, 6150, \dodoi{10.1093/mnras/stac2900}

\bibitem[{{Sekiguchi} {et~al.}(1990){Sekiguchi}, {Whitelock}, {Feast},
  {Barrett}, {Caldwell}, {Carter}, {Catchpole}, {Laing}, {Laney}, {Marang}, \&
  {van Wyck}}]{1990MNRAS.246...78S}
{Sekiguchi}, K., {Whitelock}, P.~A., {Feast}, M.~W., {et~al.} 1990, \mnras,
  246, 78

\bibitem[{{Simcoe} {et~al.}(2008){Simcoe}, {Burgasser}, {Bernstein}, {Bigelow},
  {Fishner}, {Forrest}, {McMurtry}, {Pipher}, {Schechter}, \&
  {Smith}}]{2008SPIE.7014E..0US}
{Simcoe}, R.~A., {Burgasser}, A.~J., {Bernstein}, R.~A., {et~al.} 2008, in
  Society of Photo-Optical Instrumentation Engineers (SPIE) Conference Series,
  Vol. 7014, Ground-based and Airborne Instrumentation for Astronomy II, ed.
  I.~S. {McLean} \& M.~M. {Casali}, 70140U, \dodoi{10.1117/12.790414}

\bibitem[{{Skrutskie} {et~al.}(2006){Skrutskie}, {Cutri}, {Stiening},
  {Weinberg}, {Schneider}, {Carpenter}, {Beichman}, {Capps}, {Chester},
  {Elias}, {Huchra}, {Liebert}, {Lonsdale}, {Monet}, {Price}, {Seitzer},
  {Jarrett}, {Kirkpatrick}, {Gizis}, {Howard}, {Evans}, {Fowler}, {Fullmer},
  {Hurt}, {Light}, {Kopan}, {Marsh}, {McCallon}, {Tam}, {Van Dyk}, \&
  {Wheelock}}]{2006AJ....131.1163S}
{Skrutskie}, M.~F., {Cutri}, R.~M., {Stiening}, R., {et~al.} 2006, \aj, 131,
  1163, \dodoi{10.1086/498708}

\bibitem[{{Speck} {et~al.}(2000){Speck}, {Barlow}, {Sylvester}, \&
  {Hofmeister}}]{2000A&AS..146..437S}
{Speck}, A.~K., {Barlow}, M.~J., {Sylvester}, R.~J., \& {Hofmeister}, A.~M.
  2000, \aaps, 146, 437, \dodoi{10.1051/aas:2000274}

\bibitem[{{Steffen} \& {Koning}(2011)}]{2011apn5.confP..59S}
{Steffen}, W., \& {Koning}, N. 2011, in Asymmetric Planetary Nebulae 5
  Conference, P59

\bibitem[{{Tatischeff} \& {Hernanz}(2007)}]{2007ApJ...663L.101T}
{Tatischeff}, V., \& {Hernanz}, M. 2007, \apjl, 663, L101,
  \dodoi{10.1086/520049}

\bibitem[{{Tatischeff} \& {Hernanz}(2023)}]{2023arXiv230201276T}
---. 2023, arXiv e-prints, arXiv:2302.01276, \dodoi{10.48550/arXiv.2302.01276}

\bibitem[{{Walter} {et~al.}(2012){Walter}, {Battisti}, {Towers}, {Bond}, \&
  {Stringfellow}}]{2012PASP..124.1057W}
{Walter}, F.~M., {Battisti}, A., {Towers}, S.~E., {Bond}, H.~E., \&
  {Stringfellow}, G.~S. 2012, \pasp, 124, 1057, \dodoi{10.1086/668404}

\bibitem[{{Williams} {et~al.}(1991){Williams}, {Hamuy}, {Phillips},
  {Heathcote}, {Wells}, \& {Navarrete}}]{1991ApJ...376..721W}
{Williams}, R.~E., {Hamuy}, M., {Phillips}, M.~M., {et~al.} 1991, \apj, 376,
  721, \dodoi{10.1086/170319}

\bibitem[{{Woodward} {et~al.}(2020){Woodward}, {Banerjee}, \&
  {Evans}}]{2020ATel14034....1W}
{Woodward}, C.~E., {Banerjee}, D.~P.~K., \& {Evans}, A. 2020, The Astronomer's
  Telegram, 14034, 1

\bibitem[{{Wright} {et~al.}(2010){Wright}, {Eisenhardt}, {Mainzer}, {Ressler},
  {Cutri}, {Jarrett}, {Kirkpatrick}, {Padgett}, {McMillan}, {Skrutskie},
  {Stanford}, {Cohen}, {Walker}, {Mather}, {Leisawitz}, {Gautier}, {McLean},
  {Benford}, {Lonsdale}, {Blain}, {Mendez}, {Irace}, {Duval}, {Liu}, {Royer},
  {Heinrichsen}, {Howard}, {Shannon}, {Kendall}, {Walsh}, {Larsen}, {Cardon},
  {Schick}, {Schwalm}, {Abid}, {Fabinsky}, {Naes}, \&
  {Tsai}}]{2010AJ....140.1868W}
{Wright}, E.~L., {Eisenhardt}, P. R.~M., {Mainzer}, A.~K., {et~al.} 2010, \aj,
  140, 1868, \dodoi{10.1088/0004-6256/140/6/1868}

\bibitem[{{Zubko} {et~al.}(1996){Zubko}, {Mennella}, {Colangeli}, \&
  {Bussoletti}}]{1996MNRAS.282.1321Z}
{Zubko}, V.~G., {Mennella}, V., {Colangeli}, L., \& {Bussoletti}, E. 1996,
  \mnras, 282, 1321, \dodoi{10.1093/mnras/282.4.1321}

\end{thebibliography}
\bibliographystyle{aasjournal}

\end{document}